# Bilayer twisting as a mean to isolate connected flat bands in a Kagome lattice through Wigner crystallization

Jing Wu(吴静)[1,2], Yuee Xie(谢月娥)[1,2*], Mingxing Chen(陈明星)[3], Jiaren Yuan(袁加仁)[2], Xiaohong Yan(颜晓红)[2], Shengbai Zhang(张绳百)[4] and Yuanping Chen(陈元平)[1,2*]

[1]*School of Physics and Optoelectronics, Xiangtan University, Xiangtan, Hunan, 411105, China*
[2]*Faculty of Science, Jiangsu University, Zhenjiang, 212013, Jiangsu, China*
[3]*School of Physics and Electronics, Hunan Normal University, Changsha 410081, China*
[4]*Department of Physics, Applied Physics, and Astronomy Rensselaer Polytechnic Institute, Troy, New York, 12180, USA.*

## Abstract

The physics of flat band is novel and rich but difficult to access. In this regard, recently twisting of bilayer van der Waals (vdW)-bounded two-dimensional (2D) materials has attracted much attention, because the reduction of Brillouin zone will eventually lead to a diminishing kinetic energy. Alternatively, one may start with a 2D Kagome lattice, which already possesses flat bands at the Fermi level, but unfortunately these bands connect quadratically to other (dispersive) bands, leading to undesirable effects. Here, we propose, by first-principles calculation and tight-binding modeling, that the same bilayer twisting approach can be used to isolate the Kagome flat bands. As the starting kinetic energy is already vanishingly small, the interlayer vdW potential is always sufficiently large irrespective of the twisting angle. As such the electronic states in the (connected) flat bands become unstable against a spontaneous Wigner crystallization, which is expected to have interesting interplays with other flat-band phenomena such as novel superconductivity and anomalous quantum Hall effect.

*Correponding author: yueex@ujs.edu.cn and chenyp@ujs.edu.cn.

While graphene is known for its Dirac-like band structure near which electrons can move exceptionally fast with velocities on the order of $10^6$ m/s[1-3], recently a new and opposite type of electronic behavior in graphene emerges. In a twisted bilayer graphene (TBG), due to the formation a long-range ordered Moiré pattern[4-6], these fast electrons are no longer delocalized, but instead completely localized forming a Mott insulator and subsequently a superconductor [7-22]. Xie et al. proposed that such a superconducting behavior is originated from strong interactions of the electrons in the flat bands [23,24]. Shi et al. showed that there exists an intrinsic pseudo-magnetic field in the TBGs, which is tuned by the twisting angle [25]. In addition to the electronic properties, the TBG's thermal properties can also be controlled by twisting angles[26]. These findings have attracted further attention to study the TBGs. For example, the twist angle can now be precisely controlled in experiment. Raman spectroscopy and Scanning electron microscope (SEM) techniques have been applied to identify the physical properties of the TBGs [27-33]. Theoretical calculations and analyses have also been carried out. Except for the electronic "magic" angle in 2D thin films like TBGs and twisted bilayer boron nitride(TBNN)[34], there also exists photonic "magic" angle in twisted bilayer α-$MoO_3$, whose topological polaritons transition can be controlled by twist angles[35]. As a matter of fact, the study of twisted bilayers has quickly moved beyond graphene, leading to the dawn of "twistronics"[36-38].

Besides graphene, there are other two-dimensional (2D) carbon structures in the literature, such as T-[39],Pha-[40,41], and TPH-graphene[42], as well as Kagome graphene (KG) [43,44]. While starting as theoretical predictions, some of the structures have been experimentally synthesized [42]. They exhibit a rich variety of electronic properties, which are often completely different from those of graphene. For example, the KG is a carbon allotrope made of triangular carbon rings, as shown in Fig. 1(a). There exists a flat band that touches the Fermi level. A partially-occupied flat band is expected to host a range of exotic physical phenomena such as ferromagnetism[45-47], Wigner crystallization[48-51], anomalous quantum Hall effect[52,53], and superconductivity[7,54,55].

Unfortunately, however, the flat band of KG is not isolated, while isolation is a prerequisite for a number of important applications[56-58]. Instead, it contacts with other dispersive energy bands at a point in the Brillouin zone (BZ) known as the quadratic touching. To isolate the flat band from the other bands, one possible approach is to dope the system. At a precise filling of the flat band with a partial filling factor of 1/6, the band becomes isolated, leading to the formation of a Wigner crystal of electrons[59]. And the other feasible method is to build model that has flat band in nature such as fractal-like geometry[58] or organometallic frameworks[60]. At a smaller filling, superconductivity may be expected also. However, a precise control of the band filling in experiment can be cumbersome , and also the acurrately model-building. For practical applications, the isolated flat band also must be robust. Hence, question arises: how to isolate the flat band(s) in a KG without the precision doping? It is important to note that the issue with quadratic touching is not KG specific. Rather, it is a general feature of the Kagome lattice, regardless the underlying materials. Hence, how to isolate the flat band addresses a major challenge in the development and utilization of *all* Kagome structures.

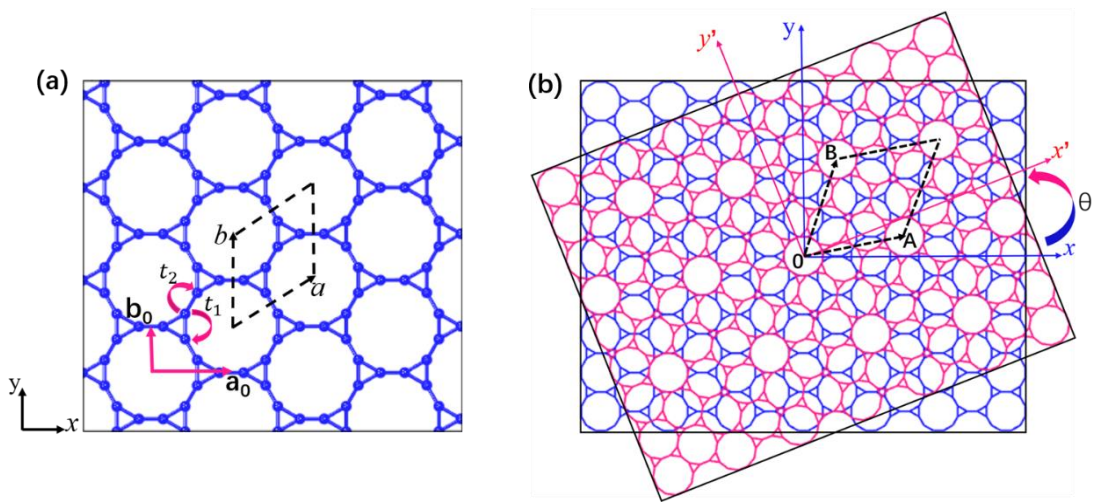

Figure 1. (a) Monolayer KG, made of triangular carbon rings. Dashed rhombus indicates the primitive cell for which the lattice constants are a and b, the basis vectors are **a₀** and **b₀**, and the intra-layer hopping parameters are $t_1$ and $t_2$. (b) A schematic view of the TBKG, in which the pink layer is twisted with respect to the blue one underneath by an angle θ around the common origin (labeled in the figure as 0). Dashed rhombus indicates the supercell (= the primitive cell of TBKG) for which the lattice constants are A and B.

In this work, we propose that an isolation of the flat band(s) can be achieved by forming a twisted bilayer, based on first-principle density functional theory (DFT) calculations. The KG has a similar crystal symmetry with graphene. Hence, depending on the twisting angle θ, various Moiré superlattices can be generated in twisted bilayer Kagome graphene (TBKGs), similar to those in graphene. Unlike the graphene, however, at a relatively large twisting angle (20° < θ < 30°) and hence in a relatively small Moiré superlattice, isolated flat bands already form. Based on the DFT results, a tight-binding (TB) model is constructed to study the isolated flat bands at smaller θ (< 10°). All the flat bands are topological nontrivial. Importantly, each isolated flat band corresponds to having a unique pattern of the Wigner crystal in real space, and vice versa. Hence, Wigner crystallization of electrons is the reason for the isolation of the flat bands. In the realm of strongly-correlated physics, the ratio between band width $t$ and interaction energy $U$ holds the key. And when the ratio is very closed to zero, in other words, the flat band is separated from others, it will have many nontrivial topogical properties[61-66]. Because the bands start to be flat in a Kagome lattice, any finite and periodic perturbation (large than the band width) is enough to cause a spontaneous crystallization, which is fulfilled and realized here by the van der Waals (vdW) interaction of the Moiré superlattice between layers.

The monolayer KG is made of motifs of triangular carbon rings, as shown in Fig. 1(a). Its primitive cell contains 6 carbon atoms, and the lattice constants are $a = b = 5.20$ Å. It has two types of bonds: one with a bond length of 1.42 Å within the triangular ring, the other with a bond length of 1.35 Å between the rings. To form the TBKG in Fig.1 (b) one rotates the top pink layer with respect to the bottom blue layer. The rotation axis is along $z$ at the center of a hexagonal honeycomb. To characterize a monolayer KG, one needs two basis vectors $\mathbf{a_0}$ and $\mathbf{b_0}$, defined as $\mathbf{a_0} = \frac{\sqrt{3}}{2}a\mathbf{i}$, $\mathbf{b_0} = \frac{1}{2}b\mathbf{j}$, where $\mathbf{i}$ and $\mathbf{j}$ are the unit vectors along $x$ and $y$, respectively [see Fig. 1(a)]. To characterize the second layer KG, one may use a rotation matrix $M[\theta] = \begin{bmatrix} cos\theta & sin\theta \\ -sin\theta & cos\theta \end{bmatrix}$, instead, where θ is the twist angle.

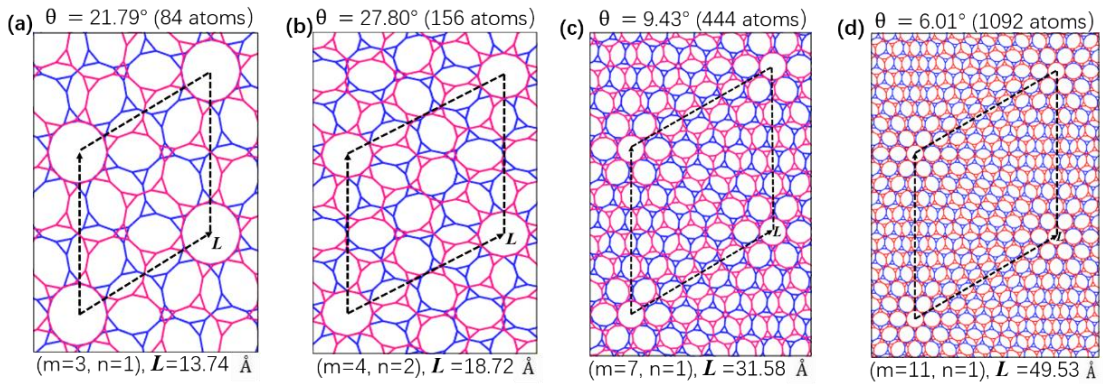

(a) θ = 21.79° (84 atoms), (m=3, n=1), $L$ =13.74 Å
(b) θ = 27.80° (156 atoms), (m=4, n=2), $L$ =18.72 Å
(c) θ = 9.43° (444 atoms), (m=7, n=1), $L$ =31.58 Å
(d) θ = 6.01° (1092 atoms), (m=11, n=1), $L$ =49.53 Å

Figure 2. Schematic illustrations of TBKGs with (a) θ = 21.79°, (b) θ = 27.80°, (c) θ = 9.43°, and (d) θ = 6.01°. Dashed rhombuses are the primitive cells for the Moiré superlattices. At the bottom of each panel, indices (m, n) are given, which are related to twisting angle θ via Eq. (1). *L* is the dimension of the primitive cell.

As θ changes, different Moiré superlattices are formed. One can also use a pair of integers (m, n) to define the twist angle θ, namely,

$$\theta = 2\tan^{-1}\left(\frac{n}{m}\right)\left(\frac{b_0}{a_0}\right). \tag{1}$$

As such, the lattice constants of the TBKG are $A = \binom{m}{n}$, $B = M[\theta]A$, respectively. The result for θ = 60° (m = n = 1) is shown in Fig. 1(b), while those for θ = 21.79° (m = 3, n = 1), 27.80° (m = 4, n = 2), 9.43° (m = 7, n = 1), and 6.01° (m = 11, n = 1) are shown in Figs. 2 (a-d), respectively. From these figures, it can be seen that the number of atoms in a Moiré unit cell grows exponentially as θ decreases, and the Moiré pattern itself becomes more pronounced. Due to the $C_6$ symmetry, structures with θ and $\theta' = 60° - \theta$ are identical. As such, one may confine angle θ in the range of 0 to 30° [67,68].

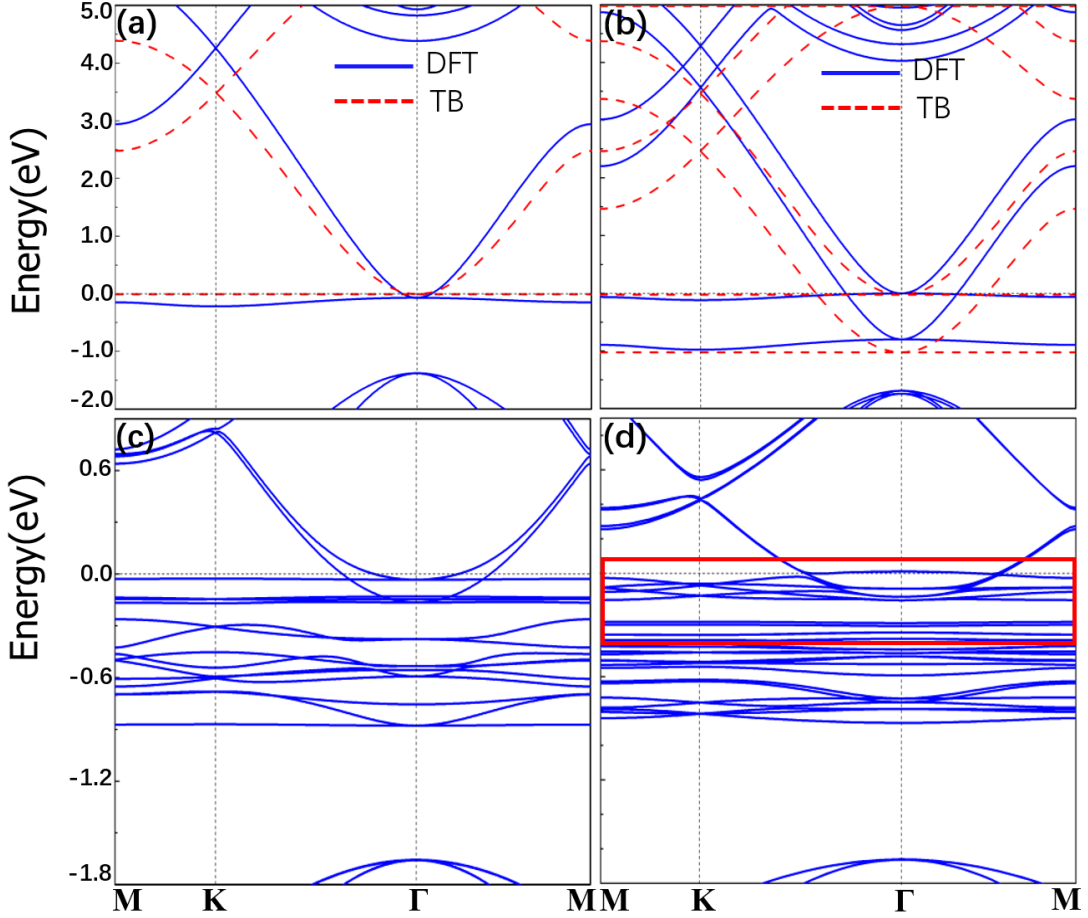

Figure 3. Band structures of (a) monolayer KG, (b) AA-stacked bilayer, (c) TBKG21.79°, and (d) TBKG27.80°. In (a) and (b), solid blue lines are the DFT results, whereas dashed red lines are the TB model. See Eq. (2). The red frame in panel (d) is enlarged as Fig. 4(a).

Our calculations were performed within DFT as implemented within the Perdew-Burke-Ernzerhof (PBE) approximation to the exchange-correlation functional. The core-valence interactions were described by the projector augmented wave (PAW) potentials, as carried out in the Vienna ab initio simulation package (VASP). Plane waves with a kinetic energy cutoff of 500 eV were used as the basis set. The Monkhorst-Pack scheme was used to sample the BZ integration. For TBKG with $\theta$ = 21.79° (TBKG21.79°) and 27.80° (TBKG27.80°), a 3×3×1 k-point mesh was used. The atomic positions were fully relaxed by the conjugate gradient method. The energy and force convergence criteria were $10^{-5}$ eV and $10^{-3}$ eV/Å, respectively. To avoid interaction between adjacent layers, a vacuum slab of 18-Å thick was used to separate the bilayers. The Grimme-D3 correction was used to account for van der Waals interactions[69,70].

Figures 3(a-b) show the band structures of monolayer KG and bilayer KG with the AA stacking, respectively. In the monolayer case, there is a flat band just below the Fermi level, which contacts quadratically with another band at Γ point of the BZ [see Fig. 3(a)]. As shown previously, upon a partial hole doping, ferromagnetism and Wigner crystallization can be realized. For the AA-stacked bilayer KG, on the other hand, Fig. 3(b) shows two sets of similar energy bands near the Femi level. The splitting between the two flat bands is approximately 1.0 eV, which indicates that van der Waals interactions between layers have a significant effect on the electronic structure, although their effects on the interlayer cohesion is small. Despite the large splitting, however, both flat bands remain in contact with neighboring bands.

Figures 3(c-d) show the band structures for TBKG21.79° and TBKG27.80°. For TBKG21.79°, each layer fits into a $\sqrt{7} \times \sqrt{7}$ supercell (measured by the monolayer primitive cell). Accordingly, the BZ is 7 times smaller than that of the primitive cell. In other words, the BZ in Fig. 3(c) has been folded 7 times. One may view the folding as starting from the AA-stacked bilayer KG in Fig. 3(b) (with 2 flat bands in the energy range between -1 and 0 eV). As a result, 14 bands are formed in the same energy range. Due to interlayer interactions, a number of the folded bands are no longer flat but dispersive, even though the dispersions are only within a couple tenths of one eV. There are, however, a handful of them that remain to be flat, e.g., the lowest one and the bands in the energy range between -0.2 and 0 eV. Also, due to the interactions, a number of the folded bands split at the boundary of the BZ (at which the bands fold). In spite of these changes, however, none of the remaining flat bands in Fig. 3(c) is isolated from the others.

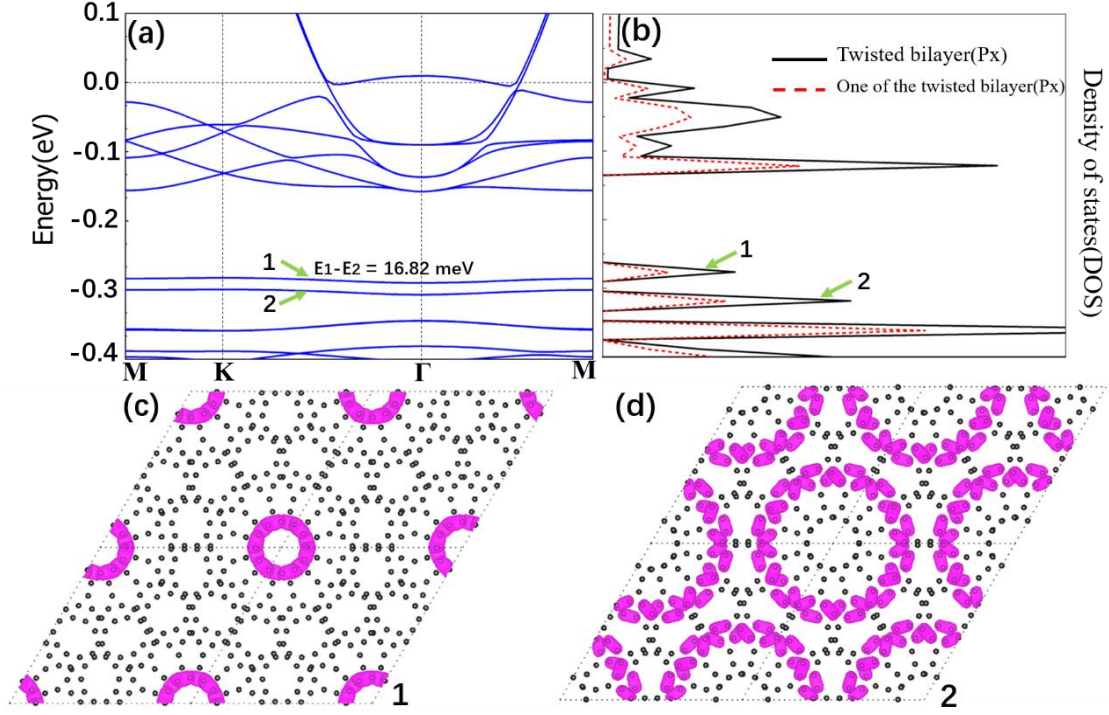

Figure 4. (a) Enlarged band structure within the red frame in Fig. 3(d). Two isolated flat bands are labeled as 1 and 2, respectively. (b) Density of states (DOS) of twisted bilayer structure. (c-d) Charge densities corresponding to band 1 and 2 in (a). Both involve enclosed circles but with different radii.

TBKG27.80° represents a different situation. Here, the TBKG contains a $\sqrt{13} \times \sqrt{13}$ supercell. As such, it has 26 bands in Fig. 3(d) in the energy range between -1 and 0 eV. While the overall features of the folded flat bands are similar to those for TBKG21.79°, there are important differences: noticeably the flat bands in the energy range between -0.3 and -0.25 eV are not only flat but also completely separated from the rest of the bands. To have a closer examination, we enlarge the band structure in the red box between -0.4 and 0.2 eV in Fig. 3(d) as Fig. 4(a). We see that there are two isolated flat bands at about -0.3 eV, labeled as band 1 and band 2, respectively. The band gap between the two is approximately 17 meV, while their band widths are only about 7 meV. There is another two isolated flat band at about -0.35 eV which constitutes, however, a set of nearly doubly degenerate states.

As mentioned earlier, a Wigner crystallization can lead to an isolated flat band. However, such a crystallization does not spontaneously happen unless the relevant flat band is at the Fermi level and can be doped controllably. In contrast, the flat bands here are not exactly at the Fermi level, although for application purposes being at the Fermi level can be advantageous. Fractional doping is also un-attempted here. The results in Fig. 4(a) thus suggest that fractional doping is not the only way to produce a Wigner crystal. Interactions between the bilayers here, despite being relatively weak, can also be an effective means to produce the Wigner crystal. To show that indeed the formation of isolated flat bands in Fig. 3(d) is a direct result of the Wigner crystallization, we plot in Fig. 4(b) the charge densities corresponding to the two flat bands, namely, band 1 and band 2. In both cases, as a hallmark of the Wigner crystallization (driven by repulsion between electrons), the charges confine themselves inside isolated rings, which in turn form a regular triangular lattice. It happens regardless the atomic structure and symmetry of the crystal. We therefore conclude that bilayer twisting is an effective way to produce isolated flat bands via a spontaneous Wigner crystallization. Here, we stress the word "spontaneous" because the isolation of the flat bands takes place in a twisted bilayer even without any atomic relaxation.

Before moving on, we would like to point out that for TBKG21.79°, in theory a spontaneous Wigner crystallization can also take place. However, it happens that the relevant flat bands are above the quadratically-touched flat band [see Fig. 3(c)]. The crossing with the dispersed band renders them not very useful. In terms of charge localization, while it can happen at k-points away from the crossing points, the band mixing at and near the crossing points ruins the possibility of having a complete Wigner crystallization.

Next, we consider TBKGs with even smaller θ (<10°). Here the large supercell size makes it impractical to perform DFT calculations. Instead, we use the tight-binding model. Only $p_z$ orbitals of the carbon atoms contribute to the electronic structure near the Fermi level. Hence, we can use the following Hamiltonian for TBKGs:

$$H = H_1 + H_2, \qquad (2)$$

where

$$H_1 = \sum_{<p,q>} \sum_\mu t_{pq} \, e^{-i\mathbf{k}\cdot \mathbf{d}_{pq}^\mu}, \qquad (3)$$

is the interlayer interaction, and

$$H_2 = \sum_{<p,p'>} \sum_v t_{pp'} \, e^{-i\mathbf{k}\cdot \mathbf{d}_{pp'}^v} + \sum_{<q,q'>} \sum_w t_{qq'} \, e^{-i\mathbf{k}\cdot \mathbf{d}_{qq'}^w}, \qquad (4)$$

is the Hamiltonian for the two monolayers. Indices *p/q* numerate the atoms in the two layers,

$\boldsymbol{d}^{\mu}_{pq}$, $\boldsymbol{d}^{v}_{pp'}$ and $\boldsymbol{d}^{w}_{qq'}$ are the relative position vectors between atomic pairs: $(p,q)$, $(p,p')$, and $(q,q')$, and $t_{pq}$, $t_{pp'}$ and $t_{qq'}$ are the corresponding hopping energies, which are defined according to the Slat-Koster semi-empirical formula[71-73], e.g.:

$$-t_{pq} = V_{pp\pi}[1 - \left(\frac{d_0}{d_{pq)}}\right)^2] + V_{pp\sigma}\left(\frac{d_0}{d_{pq}}\right)^2, \quad (5)$$

with

$$V_{pp\pi} = V^0_{pp\pi}\exp\left(-\frac{d_{pq}-c_0}{\delta}\right), \quad (6)$$

$$V_{pp\sigma} = V^0_{pp\sigma}\exp\left(-\frac{d_{pq}-d_0}{\delta}\right), \quad (7)$$

where $V^0_{pp\pi}$ is the intralayer transfer integral, $V^0_{pp\sigma}$ is the interlayer transfer integral, $d_0$ (= 3.20 Å) is the vertical distance between the layers, $c_0$ = 1.35 Å is the nearest neighbor C-C bond length, $d_{pq}$ is the interatomic distance, and $\delta$ is a decay length. For $t_{pp'}$ and $t_{qq'}$, only the two nearest-neighboring interactions $t_1$ and $t_2$ shown in Fig. 1(a), are considered. By simulating the DFT band structures of monolayer KG, AA-stacked bilayer KG, and TBKG21.79° and TBKG27.80°, we arrive at the above TB parameters. For example, $t_1$ = -3 eV, $t_2$ = -6 eV, $V^0_{pp\pi}$ = -3 eV, $V^0_{pp\sigma}$ = -0.50 eV. On the other hand, we choose $\delta$ to be 0.184 $a$ ($a$ = 5.20 Å), corresponding to having a next-nearest intralayer coupling = $0.1V^0_{pp\pi}$[74]. Judging from the comparison between DFT and TB results in Figs. 3(a-b) and those between Figs. 3(c-d) and Fig. S3 in the supplementary information (SI), it is expected that the TB model can adequately describe the TBKGs. By tuning the interlayer parameters $V^0_{pp\pi}$ and $V^0_{pp\sigma}$, it further indicates that a small interaction between layers can induce isolated flat bands (see Fig. S4 in SI).

Figure 5(a) shows the band structure for $\theta$ = 6.01°. A smaller angle typically corresponds to a larger supercell, which leads to more BZ folding and, as such, more isolated flat bands. For example, see band 1 in Fig. 5(a) and bands 2 to 5 in the zoomed-in plot in Fig. 5(b), which is over an energy range of merely 0.15 eV (between -0.78 and -0.63 eV). It is interesting to note that with a smaller twisting angle, the flat bands become so flat that the dispersion is negligible over the entire BZ. Charge densities for the five selected bands are shown in Fig. 5(c), all of which form triangular Wigner crystals. Despite that all the localized electrons are confined within closed rings, the radii of the rings can be different: e.g., bands 1 to 3 have smaller radii, while bands 4 and 5 have larger radii. In general, the tighter the radius, the less the energy dispersion. In addition to the isolated flat bands, non-isolated flat bands are also found. Figure S5 in SI depicts charge distributions for some of them. In startle contrast to the isolated flat bands, however, none of the non-isolated ones show sign of Wigner localization.

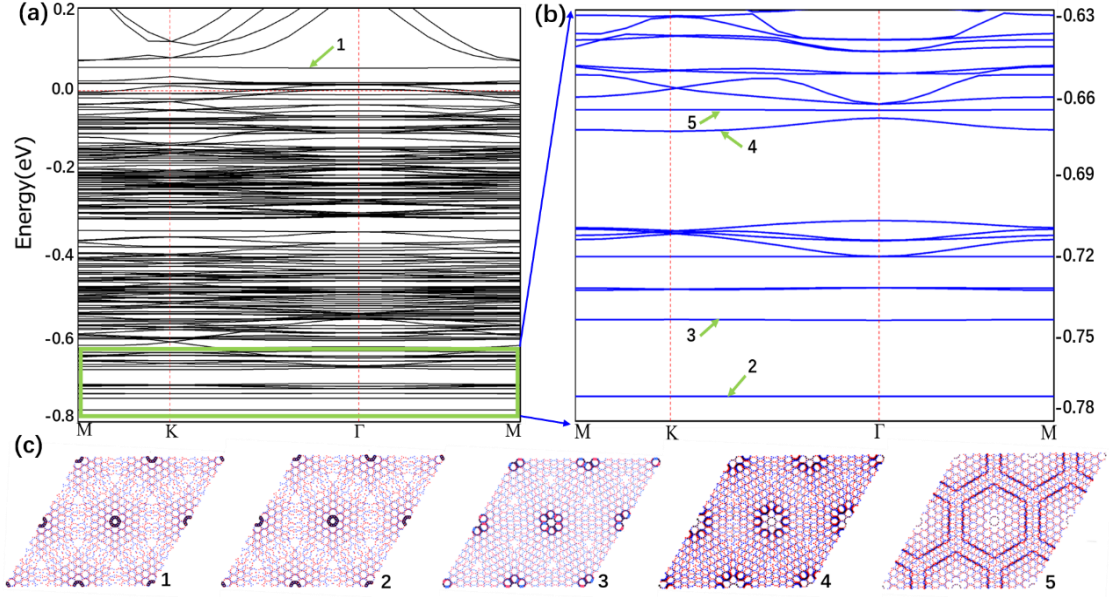

Figure 5. (a) Band structure of TBKG6.01°, based on the TB model in Eq. (2). The parameters used in the calculation are $t_1 = -3$ eV, $t_2 = -6$ eV, $V_{pp\pi}^0 = -3$ eV, $V_{pp\sigma}^0 = -0.50$ eV, and $\delta = 0.184\ a$. (b) Enlarged band structure corresponding to the green frame at the bottom of panel (a). A subset of isolated flat bands in (a) and (b) are selectively labeled as 1 to 5. (c) The corresponding charge-density patterns. All show Wigner crystallization.

Note that a twist operation here not only reduces the symmetry of the crystal but also cause a reconstruction of the atomic structure due to interlayer interaction. To see its effects on the atomic structure, Table I compares cohesive energy, interlayer spacing, and bond length for several typical stacking sequences which are AA, AA′, and AB (or equivalently BA) and twisting (given by DFT). The perfect stacking structures are shown in Fig. S6 in SI.

The cohesive energies of the stacked structures are as follows: AA′ has the lowest energy; AB is next, while AA has the highest energy. All are lower than that of monolayer KG, indicating that stacking is energetically favored. It happens that TBKG21.79° and TBKG27.80° have the same energies as the AA and AB stacking, respectively. There are two different bond lengths for monolayer KG of 1.353 and 1.424 Å. They are little change upon forming the bilayers and twisting, on average, to within (-0.0, +0.4) % for the shorter bond and (-0.2, +0.1) % for the longer bond. The general trend is that the shorter ones slightly elongate while the longer ones slightly shrink.

Table 1 Calculated layer spacing, bond length, and cohesive energy for various atomic structures.

| Structures | Layer spacing(Å) | Shorter bond length(Å) | Variation to monolayer | ratio(%) | Longer bond length(Å) | Variation to monolayer | ratio(%) | Cohesive energy (eV/atom) |
|---|---|---|---|---|---|---|---|---|
| Monolayer KG |  | 1.353 | 0 |  | 1.424 | 0 |  | -8.284 |
| AA-stacking | 3.20 | 1.356 | 0.003 | 0.22 | 1.422 | -0.002 | -0.14 | -8.305 |
| AB(BA)-stacking | 3.07 | 1.351、1.360 | -0.002、0.007 | -0.15、0.52 | 1.424、1.426 | 0、0.002 | 0、0.14 | -8.306 |
| AA'-stacking | 3.08 | 1.355 | 0.002 | 0.15 | 1.424、1.425 | 0、0.001 | 0、0.07 | -8.308 |
| TBKG21.79° | (3.15-3.21) | (1.353-1.358) | (0-0.005) | (0-0.37) | (1.420-1.424) | (-0.004-0) | (-0.28-0) | -8.305 |
| TBKG27.80° | (3.19-3.32) | (1.353-1.361) | (0-0.008) | (0-0.59) | (1.419-1.427) | (-0.005-0.003) | (-0.35-0.21) | -8.306 |

The interlayer spacing, on the other hand, is changed considerably. For example, the spacing for the AA-stacking is $d_{AA} = 3.20$ Å, which should be contrasted with those for the AA'-stacking ($d_{AA'} = 3.08$ Å) and AB-stacking ($d_{AB} = 3.07$ Å). The ratio of $\frac{d_{AA-AA'}}{d_{AA'}} = 3.9\%$, which is an order of magnitude larger than the changes in the bondlength ≤ 0.4%. A twist generally causes additional corrugation in the interlayer spacing: e.g., for TBKG21.79°, the spacing varies between 3.15 and 3.21 Å with a spread of $\Delta d = 0.06$ Å; for TBKG27.80°, it varies between 3.19 and 3.32 Å with a spread of $\Delta d = 0.13$ Å. Also, it is interesting to note that both interlayer spacings here are closer to that of AA-stacking, although energetically TBKG27.80° is the same as the AB-stacking.

Using Kagome graphene as an example, we show by DFT and TB calculations that flat band isolation can be achieved by a Moiré potential, coreated via twisting a double layer. At relatively large twisting angles (20° < θ < 30°), the potential, which is vdW in nature, is already strong enough to isolate the flat band. This is qualitatively different from the twisted bilayer graphene where the reduction of the kinetic energy to be comparable to the interlayer vdW potential requires much larger supercells and hence much smaller twisting angles. All the isolated flat bands are topological nontrivial. The physical origin for the flat band isolation is rooted in the Wigner crystallization. As a flat band possesses a number of intriguing physical properties such as ferromagnetism, anomalous quantum hall effect, and superconducting, it should be interesting to explore the interplay between these effects and Wigner crystallization.

## Acknowledgments

Work in China was supported by the National Natural Science Foundation of China (No. 11874314). SBZ was supported by U.S. DOE under Grant No. DE-SC0002623.

# References

[1] Deacon R, Chuang K-C, Nicholas R, Novoselov K, and Geim A 2007 *Phys. Rev. B.* **76** 081406

[2] de Juan F, Sturla M, and Vozmediano M A 2012 *Phys. Rev. Lett.* **108** 227205

[3] Yamoah M A, Yang W, Pop E, and Goldhaber-Gordon D 2017 *ACS nano.* **11** 9914

[4] Wang K, Qu C, Wang J, Ouyang W, Ma M, Zheng Q 2019 *ACS appl. mater.* **11** 36169

[5] Dai S, Xiang Y, and Srolovitz D J 2016 *Nano Lett.* **16** 5923

[6] Schmidt H, Rode J C, Smirnov D, and Haug R J 2014 *Nat. Commun.* **5** 5742

[7] Yankowitz M, Chen S, Polshyn H, Zhang Y, Watanabe K, Taniguchi T, Graf D, Young A F, and Dean C R J S 2019 *Science* **363** 1059

[8] Po H C, Zou L, Vishwanath A, and Senthil T 2018 *Phys. Rev. X.* **8** 031089

[9] Liu C-C, Zhang L-D, Chen W-Q, and Yang F 2018 *Phys. Rev. Lett.* **121** 217001

[10] Lian B, Wang Z, and Bernevig B A 2019 *Phys. Rev. Lett.* **122** 257002

[11] Xu C and Balents L 2018 *Phys. Rev. Lett.* **121** 087001

[12] Cao Y, Fatemi V, Fang S, Watanabe K, Taniguchi T, Kaxiras E, and Jarillo-Herrero P 2018 *Nature* **556** 43

[13] Lu X, Stepanov P, Yang W, Xie M, Aamir M A, Das I, Urgell C, Watanabe K, Taniguchi T, and Zhang G 2019 *Nature* **574** 653

[14] Kang J and Vafek O 2019 *Phys. Rev. Lett.* **122** 246401

[15] Kerelsky A, McGilly L J, Kennes D M, Xian L, Yankowitz M, Chen S, Watanabe K, Taniguchi T, Hone J, and Dean C 2019 *Nature* **572** 95

[16] Codecido E, Wang Q, Koester R, Che S, Tian H, Lv R, Tran S, Watanabe K, Taniguchi T, and Zhang F 2019 *Sci. Adv.* **5** eaaw9770

[17] Rademaker L and Mellado P 2018 *Phys. Rev. B.* **98** 235158

[18] Jiang Y, Lai X, Watanabe K, Taniguchi T, Haule K, Mao J, and Andrei E Y 2019 *Nature* **573** 91

[19] Cao Y, Fatemi V, Demir A, Fang S, Tomarken S L, Luo J Y, Sanchez-Yamagishi J D, Watanabe K, Taniguchi T, and Kaxiras E 2018 *Nature* **556** 80

[20] Cao Y, Rodan-Legrain D, Rubies-Bigorda O, Park J M, Watanabe K, Taniguchi T, and Jarillo-Herrero P 2019 *arXiv:1903.08596.*

[21] Wang W, Shi Y, Zakharov A A, Syväjärvi M, Yakimova R, Uhrberg R I, and Sun J 2018 *Nano Lett.* **18** 5862

[22] Liu Z, Li Y, and Yang Y 2019 *Chin. Phys. B.* **28** 077103

[23] Xie Y, Lian B, Jäck B, Liu X, Chiu C-L, Watanabe K, Taniguchi T, Bernevig B A, and Yazdani A 2019 *Nature* **572** 101

[24] Oh M, Wong D, Nuckolls K, Lian B, Xie Y, Watanabe K, Taniguchi T, Bernevig A, and Yazdani A 2020 *Bull. Am. Phys. Soc.*

[25] Shi H, Zhan Z, Qi Z, Huang K, van Veen E, Silva-Guillén J Á, Zhang R, Li P, Xie K, and Ji H 2020 *Nat. Commun.* **11** 1

[26] Wang M, Xie Y, and Chen Y 2017 *Chin. Phys. B.* **26** 116503

[27] Yin J, Wang H, Peng H, Tan Z, Liao L, Lin L, Sun X, Koh A L, Chen Y, and Peng H 2016 *Nat. Commun.* **7** 1

[28] Wang Y, Su Z, Wu W, Nie S, Xie N, Gong H, Guo Y, Hwan Lee J, Xing S, and Lu X 2013 *Appl. Phys. Lett.* **103** 123101

[29] Lu C-C, Lin Y-C, Liu Z, Yeh C-H, Suenaga K, and Chiu P-W 2013 *ACS nano.* **7** 2587

[30] Ta H Q, Perello D J, Duong D L, Han G H, Gorantla S, Nguyen V L, Bachmatiuk A, Rotkin S V, Lee Y H, and Rümmeli M H 2016 *Nano Lett.* **16** 6403

[31] Xing S, Wu W, Wang Y, Bao J, and Pei S-S 2013 *Chem. Phys. Lett.* **580** 62

[32] Kalbac M, Frank O, Kong J, Sanchez-Yamagishi J, Watanabe K, Taniguchi T, Jarillo-Herrero P, and Dresselhaus M S J T 2012 *J. Phys. Chem. Lett.* **3** 796

[33] Cheng Y, Huang C, Hong H, Zhao Z, and Liu K 2019 *Chin. Phys. B.* **28** 107304

[34] Xian L, Kennes D M, Tancognedejean N, Altarelli M, and Rubio A 2019 *Nano Lett.* **19** 4934

[35] Hu G, Ou Q, Si G, Wu Y, Wu J, Dai Z, Krasnok A, Mazor Y, Zhang Q, Bao Q, Qiu C-W, and Alù A 2020 *Nature* **582** 209

[36] Guinea F and Walet N R 2019 *Phys. Rev. B.* **99** 205134

[37] Carr S, Massatt D, Fang S, Cazeaux P, Luskin M, and Kaxiras E 2017 *Phys. Rev. B.* **95** 075420

[38] David A, Rakyta P, Kormányos A, and Burkard G 2019 *Phys. Rev. B.* **100** 085412

[39] Liu Y, Wang G, Huang Q, Guo L, and Chen X 2012 *Phys. Rev. Lett.* **108** 225505

[40] Wang Z, Tang C, Sachs R, Barlas Y, and Shi J 2015 *Phys. Rev. Lett.* **114** 016603

[41] Ferguson D, Searles D J, Hankel M 2017 *ACS appl. mater.* **9** 20577

[42] Fan Q, Martin-Jimenez D, Ebeling D, Krug C K, Brechmann L, Kohlmeyer C, Hilt G, Hieringer W, Schirmeisen A, and Gottfried J M 2019 *J. Am. Chem. Soc.* **141** 17713

[43] Mao J, Zhang H, Jiang Y, Pan Y, Gao M, Xiao W, and Gao H-J 2009 *J. Am. Chem. Soc.* **131** 14136

[44] Zhou M, Liu Z, Ming W, Wang Z, and Liu F 2014 *Phys. Rev. Lett.* **113** 236802

[45] You J-Y, Gu B, and Su G 2019 *Sci. Rep.* **9** 1

[46] Tasaki H 1992 *Phys. Rev. Lett.* **69** 1608

[47] Mielke A and General 1991 *J. Phys. A: Math. Gen.* **24** 3311

[48] Wigner E 1934 *Phys. Rev.* **46** 1002

[49] Padhi B, Setty C, and Phillips P W 2018 *Nano Lett.* **18** 6175

[50] Wu C, Bergman D, Balents L, and Sarma S D 2007 *Phys. Rev. Lett.* **99** 070401

[51] Bilitewski T and Moessner R 2018 *Phys. Rev. B.* **98** 235109

[52] Zhang F, Jung J, Fiete G A, Niu Q, and MacDonald A H 2011 *Phys. Rev. Lett.* **106** 156801

[53] Liu J, Ma Z, Gao J, and Dai X 2019 *Phys. Rev. X.* **9** 031021

[54] Kopnin N, Heikkilä T, and Volovik G 2011 *Phys. Rev. B.* **83** 220503

[55] Volovik G E 2018 *JETP Lett.* **107** 516

[56] Sun K, Gu Z, Katsura H, and Das Sarma S 2011 *Phys. Rev. Lett.* **106** 236803

[57] Zhang Y and Zhang C 2012 *Phys. Rev. A.* **87**

[58] Pal B and Saha K 2017 *Phys. Rev. B.* **97** 195101

[59] Chen Y, Xu S, Xie Y, Zhong C, Wu C, and Zhang S 2018 *Phys. Rev. B.* **98** 035135

[60] Liu Z, Liu F, and Wu Y 2014 *Chin. Phys. B.* **23** 077308

[61] Misumi T and Aoki H 2017 *Phys. Rev. B.* **96** 155137

[62] Tang E, Mei J W, and Wen X G 2011 *Phys. Rev. Lett.* **106** 236802

[63] Katsura H, Maruyama I, Tanaka A, and Tasaki H 2010 *EPL (Europhys. Lett.)* **91**

[64] Li X, Zhao E, and Vincent Liu W 2013 *Nat. Commun.* **4** 1523

[65] Neupert T, Santos L, Chamon C, and Mudry C 2011 *Phys. Rev. Lett.* **106** 236804

[66] Green D, Santos L, and Chamon C 2010 *Phys. Rev. B.* **82** 075104

[67] Shallcross S, Sharma S, Kandelaki E, and Pankratov O 2010 *Phys. Rev. B.* **81** 165105

[68] Zou L, Po H C, Vishwanath A, and Senthil T 2018 *Phys. Rev. B.* **98** 085435

[69] Grimme S, Antony J, Ehrlich S, and Krieg H 2010 *J. Chem. Phys.* **132** 154104

[70] Smith D G, Burns L A, Patkowski K, and Sherrill C D 2016 *J. Phys. Chem. Lett.* **7** 2197

[71] Papaconstantopoulos D and Mehl M 2003 *J. Phys.: Condens. Matter.* **15** R413

[72] Moon P and Koshino M 2012 *Phys. Rev. B.* **85** 195458

[73] Koshino M, Yuan N F, Koretsune T, Ochi M, Kuroki K, and Fu L 2018 *Phys. Rev. X.* **8** 031087

[74] Trambly de Laissardière G, Mayou D, and Magaud L 2010 *Nano Lett.* **10** 804